\documentstyle[aaspp4,11pt]{article}
\input epsf.sty

\newcommand{\bfr}{{\bf r}}
\newcommand{\bfv}{{\bf v}}
\newcommand{\half}{{\textstyle{1\over2}}}

\newcommand{\Gyr}{{\rm\,Gyr}}

\newcommand{\km}{{\rm\,km}}
\newcommand{\kms}{{\rm\,km\,s^{-1}}}
\newcommand{\kpc}{{\rm\,kpc}}

\newcommand{\pc}{{\rm\,pc}}

\newcommand{\yr}{{\rm\,yr}}

\def\spose#1{\hbox to 0pt{#1\hss}}
\newcommand{\lta}{\mathrel{\spose{\lower 3pt\hbox{$\mathchar"218$}}
     \raise 2.0pt\hbox{$\mathchar"13C$}}}
\newcommand{\gta}{\mathrel{\spose{\lower 3pt\hbox{$\mathchar"218$}}
     \raise 2.0pt\hbox{$\mathchar"13E$}}}

\begin{document}

\title{Determining the Galactic mass distribution using tidal streams 
	from globular clusters}
\author{Chigurupati Murali \& John Dubinski}
\affil{Canadian Institute for Theoretical Astrophysics \\
University of Toronto, Toronto, ON M5S 3H8, Canada}
\authoremail{murali@cita.utoronto.ca,dubinski@cita.utoronto.ca}
\begin{abstract}
We discuss how to use tidal streams from globular clusters to measure
the mass distribution of the Milky Way.  Recent proper motion
determinations for globular clusters from plate measurements and
Hipparcos astrometry provide several good candidates for Galactic mass
determinations in the intermediate halo, far above the Galactic disk,
including Pal 5, NGC 4147, NGC 5024 (M53) and NGC 5466; the remaining
Hipparcos clusters provide candidates for measurements several $\kpc$
above and below the disk. These clusters will help determine the
profile and shape of the inner halo.  To aid this effort, we present
two methods of mass determination: one, a generalization of
rotation-curve mass measurements, which gives the mass and potential
from complete position-velocity observations for stream stars; and
another using a simple $\chi^2$ estimator, which can be used when only
projected positions and radial velocities are known for stream stars.
We illustrate the use of the latter method using simulated tidal
streams from Pal 5 and find that fairly accurate mass determinations
are possible for relatively poor data sets, although current proper
motion uncertainties represent the limiting factor.  Follow-up
observations of clusters with proper motion determinations may reveal
tidal streams; obtaining radial velocity measurements can give useful
measurements of the mass distribution in the inner Galaxy.
\end{abstract}


\section{Introduction}
\label{sec:intro}
There are currently several important problems and related disputes
which depend on our understanding of the mass distribution in the
Galaxy.  In the inner Galaxy, the {\it maximum disk} controversy
revolves around determining the relative contributions of disk and
halo to the measured rotation curve (e.g. Debattista \& Sellwood 1998;
Tremaine \& Ostriker 1998).  The measurement of the rotation curve is
itself controversial (Olling \& Merrifield 1998).  At intermediate
distances in the halo, the interpretation of microlensing searches for
dark matter candidates depends fairly strongly on the shape of the
Galaxy (Alcock et al 1997).  Finally, cosmological models make strong
predictions for global halo structure (e.g. Dubinski \& Carlberg 1991;
Navarro, Frenk \& White 1997).  These results help fuel maximum disk
arguments but also predict the shape of the halo at large distances.
Resolving these issues requires significant improvements in our
understanding of the mass distribution in the Galaxy.

A variety of methods have been used to estimate the mass of the Galaxy
in different regions (see Fich \& Tremaine 1991 for a review).  In the
inner Galaxy, estimates typically rely on measurements from observable
material in the disk, including HI within the solar radius and open
clusters, OB associations and planetary nebule beyond the solar
radius.  Consequently, the mass distribution above and below the disk
is relatively poorly known (Dehnen \& Binney 1998).  In the outer
Galaxy, estimates typically rely on the dynamics of satellites and are
subject to uncertainties regarding whether individual objects are
bound to the Galaxy (e.g. Leo I) and whether or not the entire
distribution is in equilibrium given the long orbital timescales
(e.g. Little \& Tremaine 1989; Kochanek 1996).

If we could perform experiments to determine the mass, we would choose
an ensemble of test particles and study their motion in time under the
influence of the Galactic gravitational field.  Although we cannot do
this, it has been pointed out that tidal streams trace orbits in the
potential (e.g. Lynden-Bell 1982; Kuhn 1993; Johnston et al 1998), and
can therefore be used to determine the mass distribution giving rise
to that potential.  In this sense, tidal streams are analogous to
streak lines which are used to trace steady fluid flow (Batchelor
1967).

However, with the sole exception of the Magellanic stream-- whose
origin and dynamics remain controversial (Moore \& Davis 1994)--
full-fledged tidal streams remain unobserved.  Nevertheless, there is
an abundance of theoretical work which predicts the existence of tidal
streams and other substructure, either from fully disrupted, infalling
satellites (e.g. Tremaine 1993; Johnston, Hernquist \& Bolte 1996) or
from visible satellites such as globular clusters which undergo mass
loss as they orbit in the Galaxy (e.g. Gnedin \& Ostriker 1997; Murali
\& Weinberg 1997; Vesperini 1997).  In addition, the hierarchical
picture of structure formation predicts that Galactic halos should
contain a significant amount of debris from accreted substructure,
suggesting that the halo is filled with tidal streams (Johnston et al
1996).

Recent observations have finally begun to reveal traces of tidal
streams and substructure in the Galactic stellar halo.  The
Sagittarius dwarf provides an archetype for satellite accretion
(Ibata, Gilmore \& Irwin 1994) and ongoing observations are attempting
to reveal the associated stream (Mateo et al 1998).  Other
observations have revealed moving groups and phase space substructure
(e.g. Majewksi, Munn \& Hawley 1996) and extra-tidal stars surrounding
globular clusters (Grillmair et al 1995).  Given proposed astrometric
satellites, SIM and GAIA, the possibilities for using tidal streams to
probe Galactic structure appear to have multiplied dramatically
(Johnston et al 1998; Zhao et al 1999).

Given this motivation, we discuss in this paper the use of tidal
streams from globular clusters, as suggested by Grillmair (1997), to
probe the mass and potential of the Galaxy.  Theoretical work on
cluster evolution (e.g. Gnedin \& Ostriker 1997; Murali \& Weinberg
1997; Vesperini 1997) suggests that globulars with mass $M_c\lta 10^5
M_{\odot}$ and Galactocentric radius $R_g\lta 20\kpc$ will provide
excellent candidates for stream measurements because they tend to lose
mass through the combined effects of internal relaxation, tidal
heating and post-collapse heating of the core.  This suggests that
many good candidates are available.  We first summarize in
\S\ref{sec:dynamics} the dynamics of tidal streams created by mass
loss from a satellite orbiting in an external potential.  Then, in
\S\ref{sec:method}, we develop two methods for determining the
Galactic mass and potential using tidal streams.  Tests of these
methods using Pal 5 as a model cluster and of the observational
requirements are presented in \S\ref{sec:results}.  The interpretation
and importance of the results as well as further possibilities are
discussed in \S\ref{sec:discussion}.  In particular, we point out that
recent proper motion determinations from plate measurements and
Hipparcos data provide a sample of globular clusters which are good
candidates for stream observations and mass determinations.

\section{Dynamics of tidal streams}
\label{sec:dynamics}
Mass loss from globular clusters is driven by a combination of
internal relaxation, tidal heating and post-collapse heating of the
core (e.g. most recently Gnedin \& Ostriker 1997; Murali \& Weinberg
1997; Vesperini 1997).  Escaping stars reach the inner and outer
Lagrange points nearly at rest and evaporate from the system.  These
particles have slight energy offsets from the center-of-mass of the
system due to the small difference in potential energy determined by
the finite size of the satellite (Tremaine 1993; Johnston 1998).  The
cluster center-of-mass energy $E_c\sim \Phi_G(|R_c|)$, the
center-of-mass potential energy at perigalacticon; this defines a
first-order, dimensionless energy correction for escaping stars:
\begin{equation}
\delta={\Phi_G(|R_c\pm r_t|)-\Phi_G(|R_c|)\over
\Phi_G(|R_c|)}\approx \pm{r_t\over R_c}{d\ln|\Phi_G|\over d\ln R_c}
\equiv \pm\vert\chi\vert{r_t\over R_c}.
\end{equation}
The parameter $\delta=\epsilon/\Phi_G(R_c)$, where $\epsilon$ is the
energy scale defined by Johnston (1998).  The parameter
$\vert\chi\vert\leq 1$ ($\chi=1$ for a Kepler potential).  For the
typical globular clusters we will consider below, $r_t\sim 50 \pc$ and
$R_c\sim 10 \kpc$, so that $\delta\sim 0.005$; i.e. less than a 1\%
correction.  Therefore the mean motion of the stream is nearly
indistinguishable from the center-of-mass of the satellite.

Johnston (1998) finds that the absolute energies of stripped material
lies in the range $0-2\delta$ and is sharply peaked about $\delta$.
Thus the distribution of total energies lies in the range $E_{com}\lta
E_s\lta (1+2\delta)E_{com}$ (for positive $\delta$).  Therefore, the
velocity spread in the stream falls in the range $V_{com}\lta V_s\lta
\sqrt{1+2\delta}V_{com}$.  For $\delta=0.005$ and $V_{com}\sim
220\kms$, the velocity spread $\Delta v\sim 1\kms$.

\subsection{Simulated tidal streams}
The phase space coordinates of individual stream stars are determined
by the mass loss rate and fine-grained distribution of particle
positions and velocities at the Lagrange points.  Once particles are
injected into the stream, they phase mix according to the
collisionless Boltzmann equation; from the fine-grained evolution, one
can calculate the velocity and density structure along the stream
(Tremaine 1998; Helmi \& White 1999).

Here we generate streams using both simple N-body simulations and an
analytic approximation to the characteristics of the projected stream
based on the discussion of energetics given above.  In the quasistatic
evolution of globular clusters, the mass loss rate is small so the
potential remains very nearly spherical and constant over the
timescales considered here.  Therefore the N-body simulations use a
fixed, Plummer-law satellite with test particle orbits integrated
along the satellite's orbit in the Galaxy.

With N-body calculations, it is difficult to accurately reproduce
expected mass loss rates from globular clusters given the importance
of internal relaxation and its dependence on the stellar mass spectrum
as well as the possible importance of core heating in evaporating
clusters (e.g. Gnedin \& Ostriker 1997; Murali \& Weinberg 1997).
Therefore, to increase our flexibility, we adopt a simple Gaussian
approximation to simulate the projected characteristics of tidal
streams for use in the projected orbit fits discussed below.  Of
course, the dynamics of the stream are best studied through direct
orbit integration and N-body simulation: below we compare this
approximation with the results of a simulation to ensure that the
approach is reasonable.

In this approximation, we adopt a mass loss rate to specify the number
of stars in the stream and assume that the stream stars have Gaussian
distributions of: 1) orbital phases about the current phase of the
satellite; 2) line-of-sight, radial velocities about the
line-of-sight, radial velocity of the satellite at the star's phase;
3) angular offsets from the angular position of the satellite at the
star's phase.  The phase distribution is very narrow for recent mass
loss; this corresponds to a distribution clumped about the satellite.
The phase distribution is very broad for mass loss in the distant
past; this corresponds to a uniform phase distribution or a
phase-mixed stream.  The dispersion of the radial velocity
distribution is given roughly by the velocity range determined from
the energy spread.  The angular dispersion corresponds to the angular
size of the system at the star's orbital phase.

In practice, we choose phase by sampling time along the orbit since
azimuthal phase angle $w=\Omega t$, where $\Omega$ is the azimuthal
frequency of the orbit.  The Galactic latitude of the satellite at
this time is chosen as the phase variable \footnote{This is usually
unique for small angular scales.  Sometimes it may be necessary to
choose a different independent variable-- e.g. if the stream makes a
loop on the sky in $l$.}.  Then radial velocity and angular variates
are generated assuming means given by the center-of-mass coordinates
at this phase.  This provides a reasonable approximation for the
characteristics of a stream from a globular cluster.  For a larger
satellite, it is necessary to account for the offset of the stream
from the orbit of the satellite (Johnston 1998).  It is useful because
we can arbitrarily change the number of stars in the stream and their
phase distribution in order to explore observational possibilities.

We choose mass loss rates in the range
\begin{equation}
{\dot M\over M}\equiv \lambda=0.1-1.0 \times 10^{-10} \yr^{-1}.
\end{equation}
The mass loss rates imply that clusters have lost roughly $40-60\%$ of
their initial mass for fixed $\lambda$.  The average rate is
consistent with (and even somewhat lower than) recent calculations of
the evolution of relatively low mass clusters: $M_c\lta 10^5
M_{\odot}$ (Gnedin \& Ostriker 1997; Murali \& Weinberg 1997;
Johnston, Sigurdsson \& Hernquist 1998).  For convenience, we define
the parameter $\lambda_{-10}$ to have units $10^{-10}\yr^{-1}$ so that
$0.1\lta\lambda_{-10}\lta 1$.  We assume a mean stellar mass $\langle
m\rangle=0.5 M_{\odot}$ for the cluster to define the number of stars
in the stream.  The importance of the stellar content of the stream is
discussed below.

\section{Using tidal streams for mass and potential determinations}
\label{sec:method}
We develop two methods which can be used to determine the mass and
potential of the Galaxy using tidal streams.  The first method assumes
that a tidal stream very nearly follows a streamline in the Galactic
potential, which is a very good approximation for a globular cluster.
With complete phase space data for stream stars, we can determine the
Galactic mass and potential directly from the observations without any
modeling.  This approach is a generalization of rotation-curve mass
measurements to non-circular orbits.  The upcoming space astrometry
missions, SIM and GAIA, promise to give phase-space coordinates for
nearby clusters ($d\lta 5\kpc$) and their tidal streams which will be
sufficiently accurate to use this method effectively.

The second method involves fitting a model stream curve to stream data
where we know the full position and velocity information for the
cluster but only know projected positions and radial velocities for
stream stars.  Here we are motivated by the possibility of combining
the results of various plate measurement programs (e.g. Dinescu et al
1999) and the Hipparcos results on globular cluster proper motions
(Odenkirchen et al 1997) with current ground-based observational
capabilities.  We discuss the possibilities in more detail below.

\subsection{The streamline approximation}
Tidal streams approximately trace the orbit of their parent satellite
in the Galaxy.  With full phase space information, Lynden-Bell (1982),
Kuhn (1993) and Johnston et al (1998) point out that we can
approximately measure the potential difference along the stream by
measuring the kinetic energy difference between different positions
since the force field is conservative.  For globular clusters, the
streamline approximation should be quite accurate since the energy
spread in the stream is quite small.

For completeness, we present the basic derivation of the equation of
energy conservation (Bernoulli's equation) for a streamline starting
with the equation of motion.  The acceleration of a particle in a
gravitational potential is given by Newton's law:
\begin{equation}
{d\bfv\over dt}=-\nabla\Phi(\bfr).
\end{equation}
Because the orbit is parameterized by time, this equation provides
little information about the potential of the Galaxy.  However,
rewriting it in Lagrangian form (assuming a static potential),
\begin{equation}
{d\bfv\over dt}={\partial \bfv\over \partial \bfr}\cdot{d\bfr\over dt}=
{\partial\bfv\over \partial \bfr}\cdot\bfv=-\nabla\Phi(\bfr),
\label{eq:path_acc}
\end{equation}
we parameterize the motion in terms of the path of the particle.  The
acceleration of the particle may therefore be determined from the
velocity history of the particle along the path.

While the path of an individual star is unknown, a tidal stream traces
the path of a hypothetical particle in the Galactic potential.  Since
we can, in principle, determine positions and velocities of material
along the stream, we can calculate the acceleration at a point using
equation (\ref{eq:path_acc}).

Equation (\ref{eq:path_acc}) is equivalent to the equation of motion
for an element of an incompressible fluid in a static potential
(Batchelor 1967).  In this context, we may view a tidal stream as the
manifestation of a streamline.  By observing the velocity field along
the streamline, we can determine the gravitational potential which
produces it.

With no symmetry assumption, we can obtain general expressions which
define the potential along the path.  Taking the line integral of
equation (\ref{eq:path_acc}) along the path, we determine the
potential difference between two points along the curve
\begin{equation}
\Phi(\bfr_1)-\Phi(\bfr_0)\equiv\Delta\Phi_{01}=-\int_{\bfr_0}^{\bfr_1}
d\bfr\cdot{\partial \bfv\over \partial \bfr}\cdot{\bfv}.
\end{equation}
Since the velocity field is irrotational (except for the possibility
of contamination by binaries and spin in the satellite itself), this
can be written
\begin{equation}
\Delta\Phi_{01}=-\int d\bfr
{dv^2/2\over d\bfr}=\half[v^2(\bfr_0)-v^2(\bfr_1)].
\end{equation}
This is simply a statement of energy conservation (Bernoulli's
equation) but makes the point that a measurement of the difference in
the kinetic energy of two points along the stream is equivalent to a
measurement of the potential difference or work done between the two
points.  Thus, if we can measure an ensemble of tidal streams, we will
be able to reconstruct the Galactic potential in a fairly unbiased
manner.

Now, assuming a spherically symmetric potential, we easily obtain the
mass from the velocity gradient:
\begin{equation}
M(r)=-{r^2\over G}\bfv\cdot{\partial \bfv\over \partial \bfr}\cdot{\hat r}.
\label{eq:mass1}
\end{equation}
For measurement purposes, the equation of energy conservation provides
a more convenient way to estimate the spherical mass.  Rewriting
$\bfv\partial\bfv/\partial\bfr$ assuming zero vorticity, we find
\begin{equation}
M(r)=-{r^2\over G}{\partial v^2/2\over \partial \bfr}\cdot{\hat r}.
\end{equation}
For a circular orbit, we recover the usual formula for rotation curve
measurements:
\begin{equation}
M(r)={rv_c^2\over G},
\end{equation}
where $v_c$ is the circular rotation velocity.  

By considering only the radial force component in equation
(\ref{eq:mass1}), we have ignored the information provided in the
other directions.  In general, accurate determination of the velocity
field as a function of 3-dimensional position along the stream
directly gives the 3 components of the gravitational acceleration.
This, in turn, provides information on the asphericity of the mass
distribution.  In fact, the non-radial components of the acceleration
can be determined most accurately since they only depend on
differences of angular coordinates, which are determined very
accurately.  For complete generality, we can simply take the
divergence of equation (\ref{eq:path_acc}) and obtain a dynamical form
of Gauss' law:
\begin{equation}
\nabla\cdot{\partial\bfv\over \partial \bfr}\cdot\bfv=
	{1\over 2}\nabla^2 v^2=-\nabla^2\Phi(\bfr)=-4\pi G\rho.
\end{equation}
This, of course, has the disadvantage that second derivatives are
required so that the data must be very accurate.  Thus it does not
appear to be of immediate practical use.

\subsection{Fitting the projected stream}
\label{sec:statistical}
Given the difficulty of obtaining a complete set of data, we suggest a
statistical approach to local mass determinations.  What we describe
is a procedure for fitting a projected tidal stream to the
observational data using a $\chi^2$ estimator.  This is similar to the
method described by Johnston et al (1998) but only requires projected
positions and radial velocities for material along the stream and does
not depend on the structure of the satellite.

Suppose we have a satellite with determined position and velocity, and
we observe an associated stellar stream for which we can measure only
projected positions and radial velocities of individual stars.  Given
a model mass distribution defined by some set of parameters
${\bf\theta}$ and the satellite position and velocity, we can
integrate the satellite trajectory forward and backward in phase for
each set ${\bf\theta}$ and find the trajectory which best fits the
stream data.  It is straightforward to use a $\chi^2$ estimator,
choosing Galactic longitude $\ell$ as the independent, phase variable
and Galactic latitude $b$ and radial velocity $v_r$ as dependent
variables:
\begin{equation}
\chi^2=\sum_i^N\biggl({b_i-b(\ell_i|{\bf\theta})\over \sigma_{b,i}}
	\biggr)^2+\sum_i^N\biggl({v_{r,i}-v_r(\ell_i|{\bf\theta})\over
	\sigma_{v,i}} \biggr)^2;
\label{eq:chisq}
\end{equation}
as usual, this defines the logarithm of the joint
probability\footnote{ Strictly speaking, this is the joint probability
{\it density}} of measuring $b_i$ and $v_{r,i}$ at $\ell_i$, given the
model.  Note that it is straightforward to generalize this procedure
when more information is available.

The dispersions $\sigma_{b,i}$ and $\sigma_{v,i}$ include
contributions from the width and velocity dispersion in the stream and
uncertainties in the observations.  Observational uncertainties are
negligibly small for $\sigma_{b,i}$ but potentially dominate
$\sigma_{v,i}$.  In addition to the physical dispersion in position
and velocity, the angular length of the observed stream and the number
of stars observed strongly determine the quality of the mass
determination.  We consider the role of these factors below.

\subsubsection{Including proper motion uncertainties}
\label{sec:pm_sig}
In presenting the $\chi^2$-estimator, we have implicitly assumed that
all available data are determined to high precision.  In general, this
may not be the case: indeed, we currently face fairly broad
uncertainties in proper motion measurements for individual clusters.
It is, nevertheless, straightforward to include measurement
uncertainties in the curve-fitting procedure from a Bayesian point of
view.

For the specific example of proper motion uncertainties, we can add
two parameters to our model: namely $\mu_{\alpha}$ and $\mu_{\delta}$,
the proper motions measured with respect to right ascension and
declination, respectively.  Since we have estimates and uncertainties
for these two quantities, by Bayes' theorem (e.g. Martin 1971) the
probability of the model given the data becomes
\begin{equation}
P({\bf\theta}')\propto \prod_i P_i({\bf\theta}') 
	P(\mu_{\alpha})P(\mu_{\delta});
\end{equation}
in other words, the relative probability of any set of parameters
${\bf\theta}'$, given the data, is the joint probability of the data
given the model, $\prod_i P_i({\bf\theta}')=\exp(-\chi^2/2)$
multiplied by the prior probabilities of the proper motions,
$P(\mu)=\exp[-(\mu-\mu_0)^2/ \sigma_{\mu}^2]$, where $\mu$ denotes
either $mu_{\alpha}$ or $\mu_{\delta}$ and $\mu_0$ the respective
mean.  To return to the original set of parameters ${\bf\theta}$, we
project over $\mu_{\alpha}$ and $\mu_{\delta}$:
\begin{equation}
P({\bf\theta})\propto \int d\mu_{\alpha} d\mu_{\delta} P({\bf\theta}').
\label{eq:P_theta}
\end{equation}
We examine the influence of proper motion uncertainties in \S4.2.1.

\section{Mass estimates in spherical potentials}
\label{sec:results}
For illustrative purposes, it is simplest to adopt spherical,
scale-free mass models for the Galaxy:
\begin{equation}
M(<r)=M_0 \biggl ({r\over r_0} \biggr )^{\alpha},
\end{equation}
where $M_0$ is the mass interior to some assumed radius $r_0$ and
$\alpha$ gives the slope of the cumulative mass distribution.  The
model has two parameters $M_0$ and $\alpha$.

In the discussion, we point out several candidate clusters to which we
can apply the spherical mass estimator discussed here.  Among these
are Pal 5: for definiteness in specifying satellite initial
conditions, we adopt the best estimates for the space motion for Pal
5.  Pal 5 is a low mass globular cluster, $M_c\sim 10^4 M_{\odot}$, at
high Galactic latitude and distance $d=20\kpc$ on the far side of the
Galactic center with a fairly elongated orbit.  There are two
conflicting proper motion determinations (Cudworth \& Majewski 1993;
Scholz et al 1998) but both indicate that this cluster has just passed
apogalacticon and is not associated with the Sgr dwarf galaxy as had
been suggested by Ibata et al (1997).

Here we adopt the more recent proper motion
$(\mu_{\alpha},\mu_{\delta})=(-1.0\pm0.3,-2.7\pm0.4) {\rm mas/yr}$ and
3-dimensional velocity determined by Scholz et al (1998; see their
Table 2) and integrate the orbit in a spherical potential as defined
above with $M_0=2\times 10^{11} M_{\odot}$, $r_0=20\kpc$ and
$\alpha=1$ (singular isothermal sphere).  The orbital period is
$2.4\times 10^8 \yr$.  We convert to galactocentric quantities using
the formulae given in Johnson \& Soderblom (1987), assuming a solar
distance $R_{\odot}=8.5\kpc$, a rotation velocity of $220 {\rm km/s}$
and the `basic solar motion' (Mihalas \& Binney 1982).  Although the
Galactic mass distribution along the entire orbit is not spherical, we
are only interested in the prospects for mass determination from
observations on relatively small angular scales: $\lta 10^o$.  The
stream material over this angular range remains high above the
Galactic plane and extends roughly $5\kpc$ in length so that
deviations from spherical symmetry will be small.

\subsection{With full phase space information}
Here we generate an example stream using an N-body simulation to see
if the mass is recovered correctly.  Figure 1 shows that we do obtain
the correct mass.  In this case, the satellite orbit was started 5
radial periods in the past (approximately $1.2\Gyr$) so that there
have been 5 perigalactic passages.  Most of the mass loss occurred in
the first passage, so that material has had time to drift away from
the satellite, spreading over an angular extent
$\vert\Delta\Theta\vert<3^o$. This is advantageous because, near the
satellite, the stream stars have more complicated dynamics since they
are still far their asymptotic energy distribution; mass
determinations using particles near the satellite will be biased.
Distances greater than twice the tidal radius should be adequate to
ensure the proper behavior.

\begin{figure}
\plotone{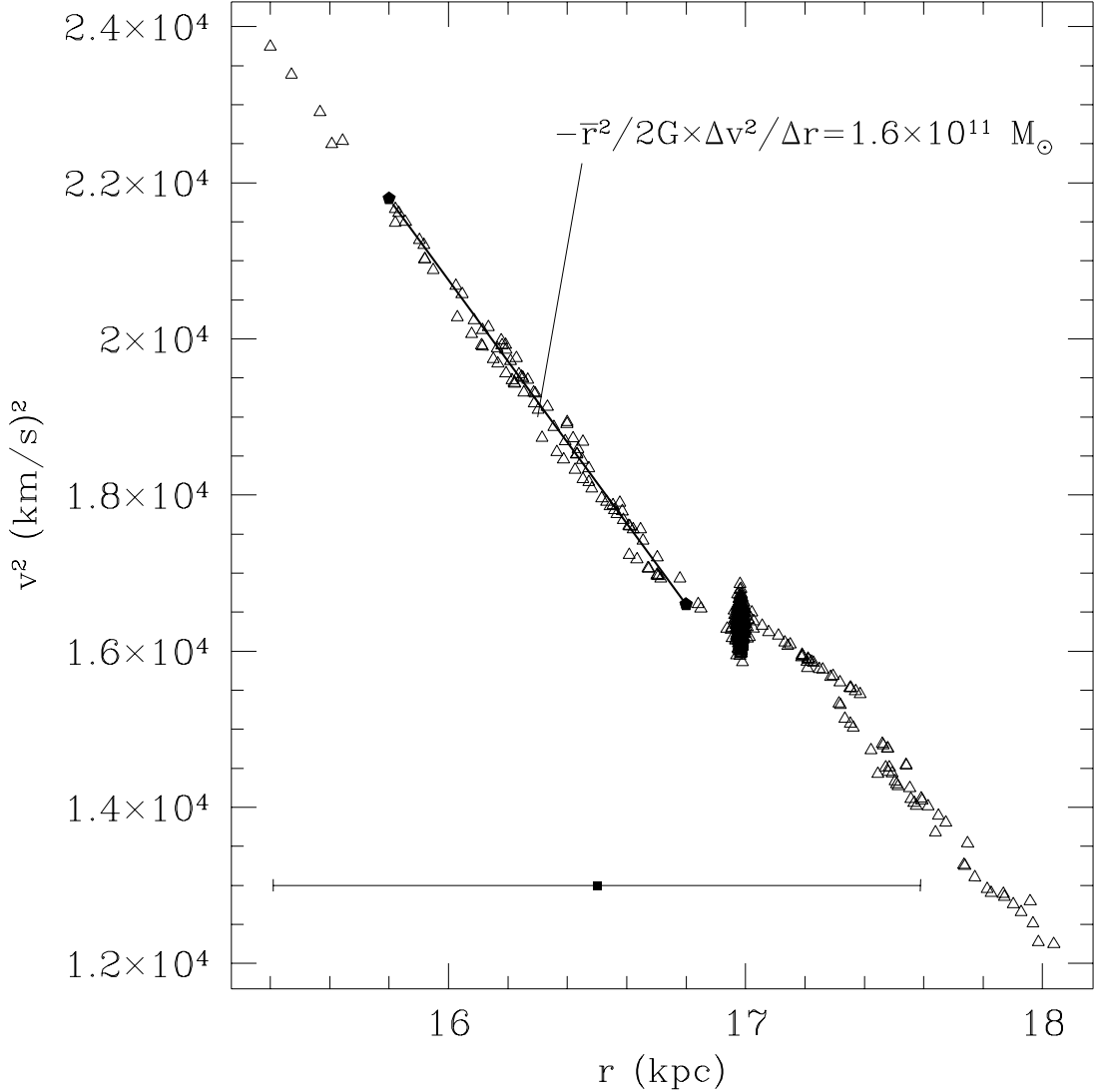}
\caption{A plot of the total square velocity vs. galactocentric radius
for a simulated cluster when full information is available for the
stream.  The plot gives the mass directly.  Note that the leading
(left) and trailing streams have energy offsets relative to the
satellite but a mass determination on either side gives the correct
mass.  The stream has an angular extent $\vert\Delta\Theta\vert<3^o$
about the satellite.  The error bar shows the $\pm 1-\sigma$
uncertainty in the SIM and GAIA distance determinations for a
satellite at $d\sim 20\kpc$.  Here uncertainties dominate the
measurement.  The method is effective at smaller distances since the
uncertainties shrink quadratically with distance while the length of
the stream drops linearly for fixed angular size.}
\end{figure}

In general, measuring the potential difference between two points
along a stream is easiest because we need only compute the kinetic
energy difference.  The mass determination is somewhat more difficult
because we must measure velocities at neighboring points and then take
differences between these two points to evaluate the derivative.
Clearly this can be sensitive to noise.  Here, $v^2$ happens to have
an approximately linear dependence on $r$ so determining the slope is
easy.  In general, the relationship will be more complicated since
$\Delta v^2\propto \Delta\Phi$.  One approach would be to fit a smooth
curve to the data and use that to extract physical quantities.

Figure 1 shows that at the distance of Pal 5, the uncertainties in
distance measurements for SIM and GAIA will dominate since
$\sigma_d\approx(d/20\kpc)^2 1.6\kpc$.  Thus, although absolute
distances are known with 10\% accuracy, the relative distances of
stream stars can be highly uncertain.  Since the uncertainties
decrease quadratically with distance, the uncertainties in relative
distance between stream stars decrease linearly with distance for
fixed angular size.  This suggests that the method can be best used
for fairly nearby clusters, $d\lta 5\kpc$, to determine the 3
components of the gravitational acceleration near the disk.

\subsection{In projection}
For convenience, we use the Gaussian approximation described above to
generate realizations of streams in projection.  As a simple check, we
compare the results of an N-body simulation with a stream generated
using this procedure.  The simulation was started 3 radial periods in
the past so that the satellite has had 3 perigalactic passages.  The
realization has phase dispersion $\sigma_t=5\times 10^6\yr$,
line-of-sight, radial velocity dispersion $\sigma_{V_r}=1\kms$ and
latitude disperision $\sigma_b=10'$.  The phase dispersion $\sigma_t$
is chosen simply by inspection of the N-body simulation.  The others
are defined by the characteristics of Pal 5.  Figure 2 shows that the
agreement is reasonable.

\begin{figure}
\plotone{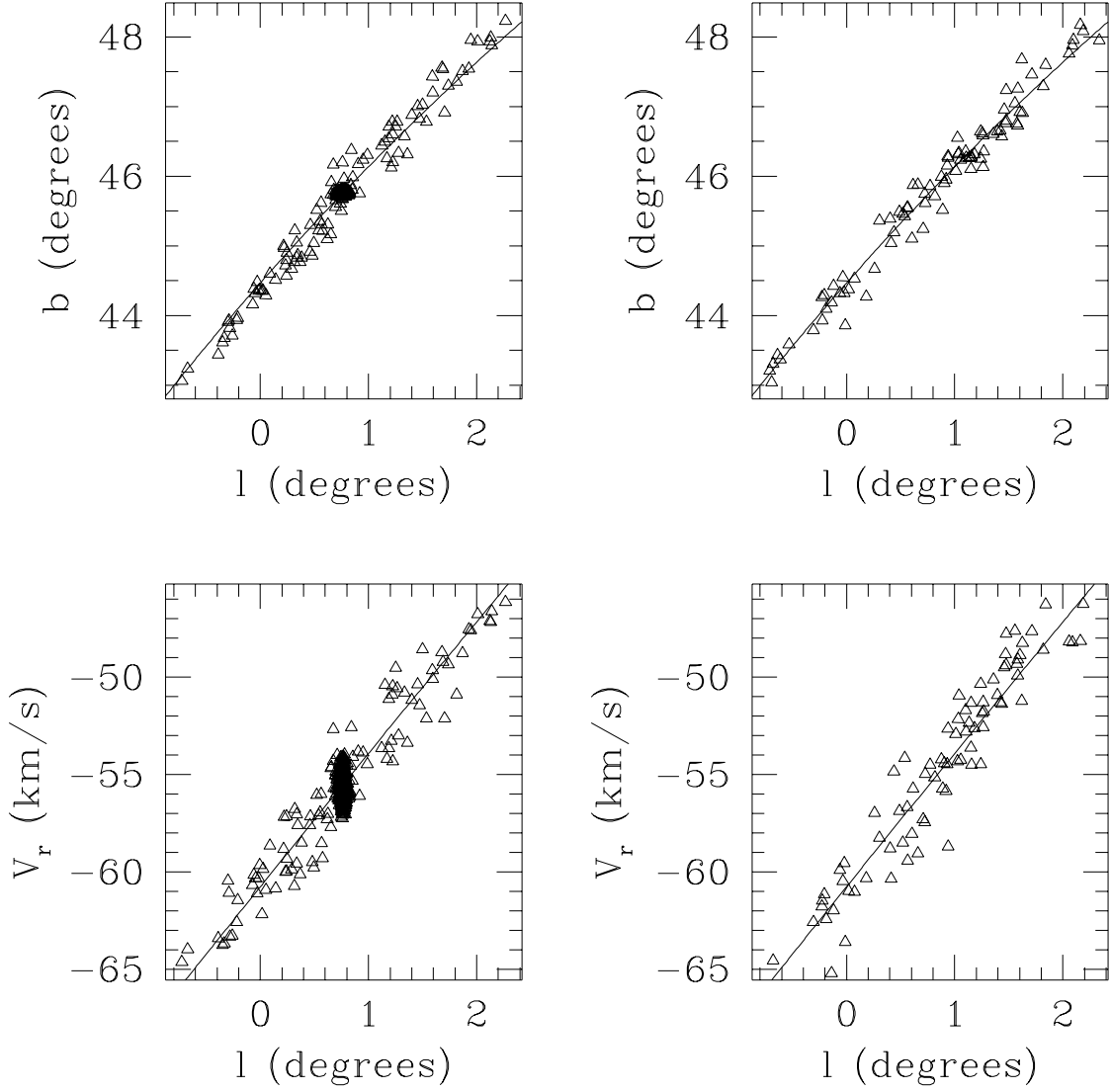}
\caption{Comparison of projected characteristics of N-body tidal
stream and Gaussian realizations.  The left panels show the $b$
vs. $\ell$ dependence (top) and $V_r$ vs. $\ell$ (bottom) dependence
for the N-body stream.  The right panels show the same dependences for
the approximation.  The solid lines in each panel show the projection
of the satellite orbit in $b$ vs. $\ell$ and $V_r$ vs. $\ell$.  The
agreement is reasonable.}
\end{figure}

Our view of a tidal stream is determined by the mass loss history of
the satellite as well as the angular extent of the observations.  If
little mass loss has occurred recently, then material will be well
mixed and more difficult to detect near the satellite.  As an example,
Figure 3 shows the observed characteristics of a stream with 56, fully
phase-mixed stars in an angular range $\vert\Delta\Theta\vert<10^o$.

This provides somewhat of a worst-case scenario because there has been
no recent mass loss and because there are very few stars in the
stream.  Nevertheless, although the orbits have indistinguishable
spatial projections for all values of $M_0$ on this scale, the radial
velocies provide a strong discriminant.  This is not suprising since,
physically, the measured mass depends sensitively on the velocities:
$M\propto v^2$.

\begin{figure}
\plotone{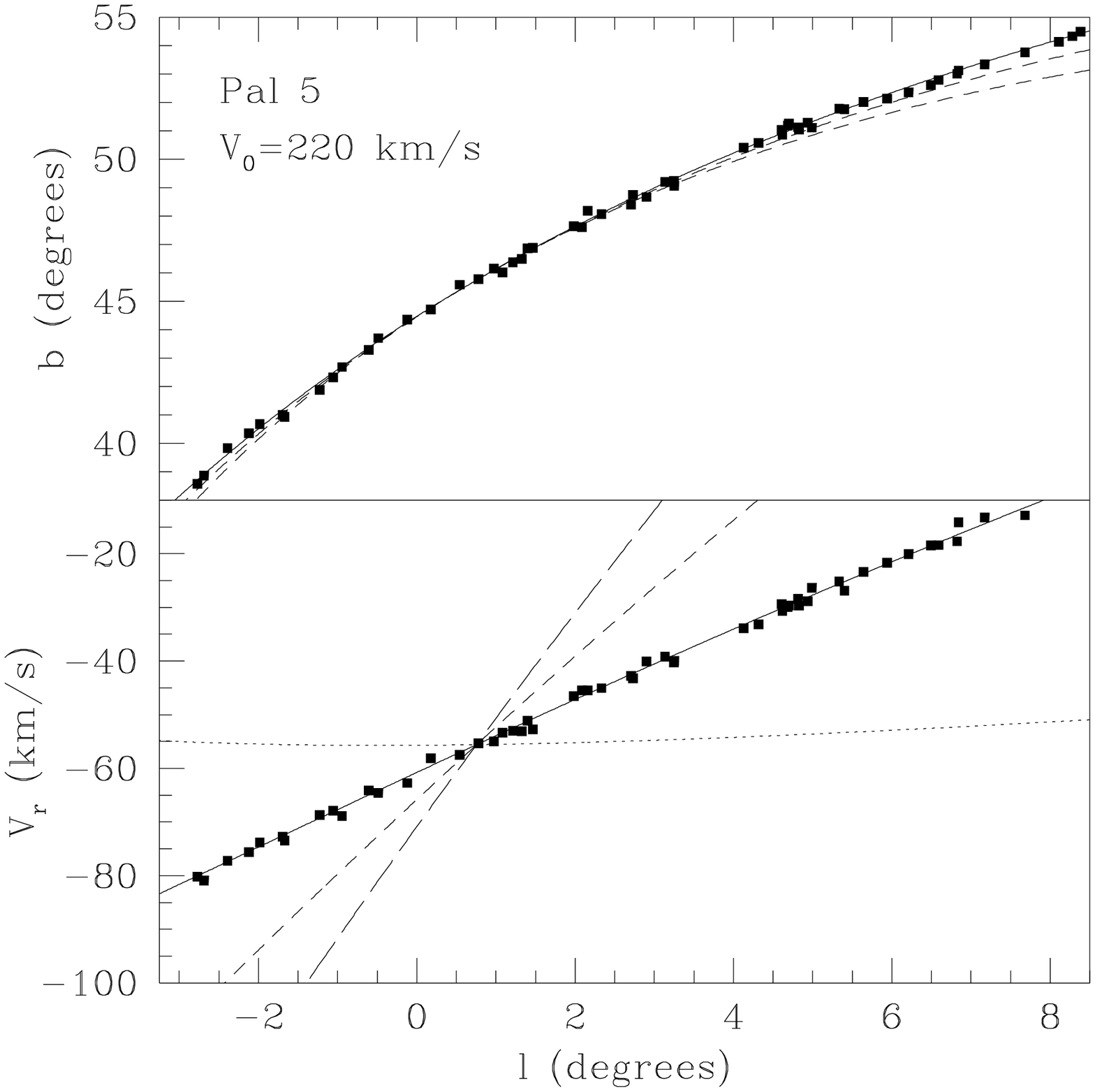}
\caption{Observed characteristics of phase-mixed tidal stream with
$\vert\Delta\Theta\vert<10^o$.  There are 56, fully phase-mixed stars
in the stream. The cluster sits at the intersection of the curves.
The top panel shows $b$ versus $l$ for Pal 5 initial conditions and
$M_0=1$ (dotted), $2$ (solid; actual value), $3$ (short dashed) and
$4$ (long dashed) $\times 10^{11} M_{\odot}$.  The bottom panel shows
the heliocentric radial velocity versus $l$ for the stream and each of
the different orbits.  The radial velocity profile provides a strong
discriminant for the mass normalizations.}
\end{figure}

To quantify this statistically , we generate fits to streams using the
orbit estimator, equation (\ref{eq:chisq}).  Figure 4 shows the
confidence intervals in the $\alpha-M_0$ parameter space for the
indicated particle number and $\Delta\Theta$ with fully phase-mixed
debris.  These results suggest that under extremely poor conditions,
the mass determination is highly uncertain; under somewhat better
conditions it is possible to constrain the mass and mass distribution
with several \% accuracy.

\begin{figure}
\plotone{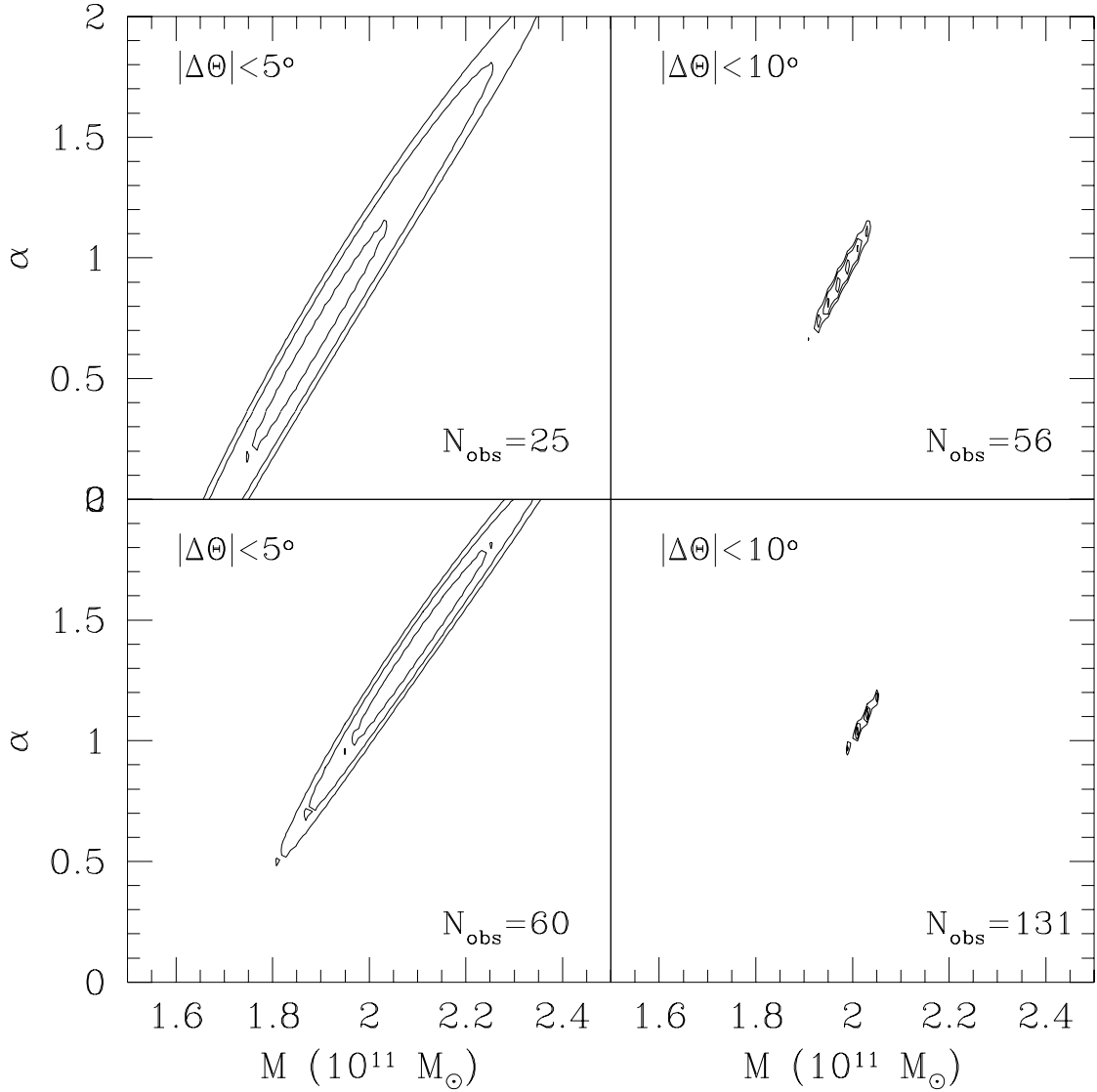}
\caption{Confidence intervals in $\alpha$ and $M_0$ for different
values of $\Delta\Theta$ and $N_{obs}$ in phase-mixed streams.
Contours show the 99\%, 95\% and 68\% confidence contours. The
upper-left gives the worst-case scenario: very small mass loss in the
distant past and a small observed angular scale.  The lower-right
gives a much better scenario: a factor of 2 increase in mass loss and
a factor of 2 increase in angular extent of observations.  In the
first case, the mass and mass distribution are both highly uncertain,
In the latter case, the confidences improve significantly: both $M_0$
and $\alpha$ can be determined to within several \%.  Comparing the
upper-right and lower-left panels suggests that larger angular extent
improves the mass determination more than does increased sample size.}
\end{figure}

Assuming fully phase-mixed debris is equivalent to assuming no recent
mass loss.  As another example, we assume a low rate of mass loss
$\lambda_{-10}=0.125$ over the last 10 perigalactic passages (roughly
$2.5\Gyr$) so that debris still clusters near the satellite and has a
smaller angular extent.  Figure 5 shows the appearance of the stream
in projection at the present time over an interval $\vert
\Delta\Theta\vert<10^o$.

\begin{figure}
\plotone{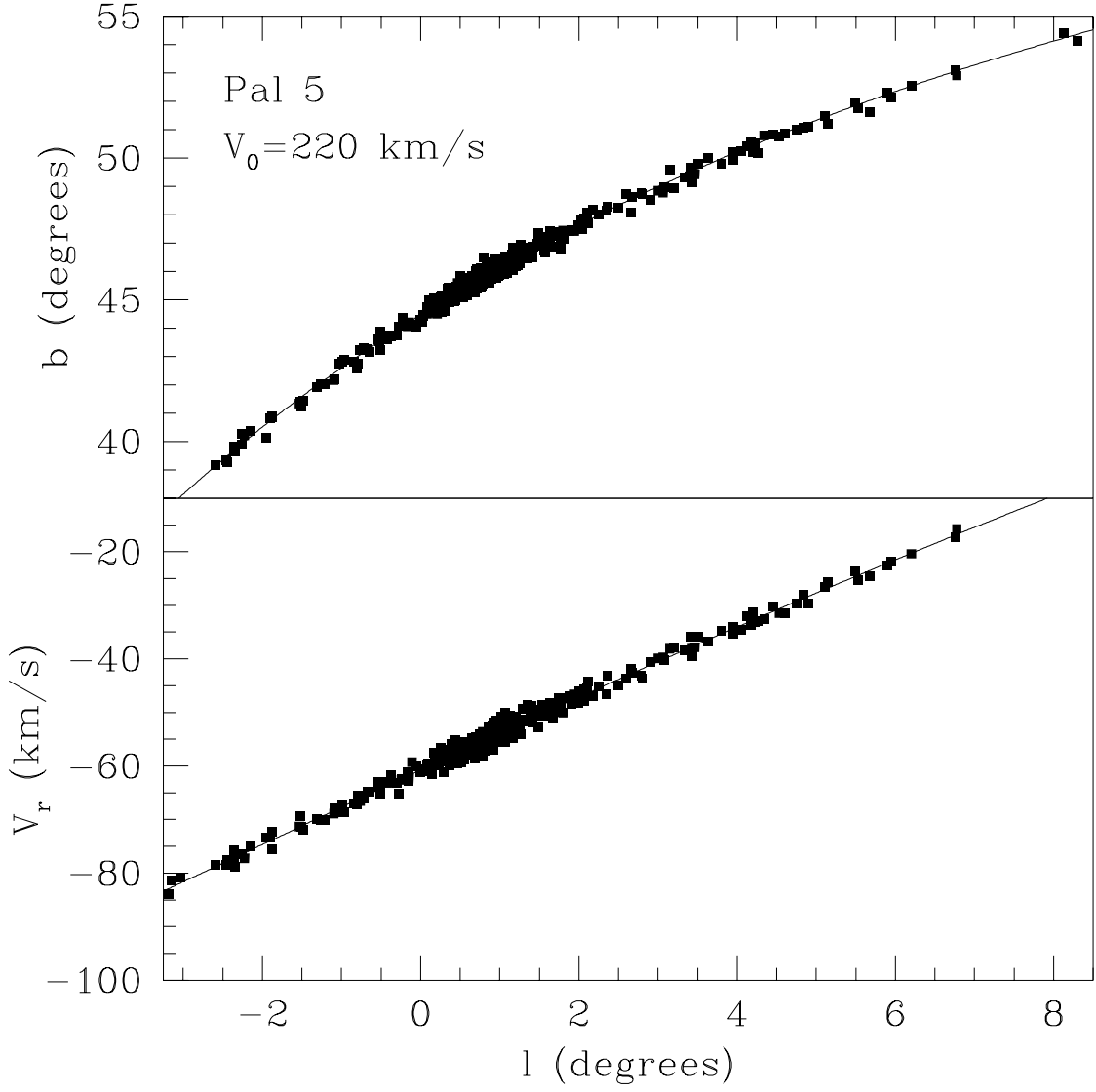}
\caption{Observed characteristics of clumped tidal stream with
$\lambda_{-10}=0.125$ and $\vert\Delta\Theta\vert<10^o$.  There are
487 stars in the stream. The top panel shows $b$ versus $l$ for Pal 5
initial conditions $2\times 10^{11} M_{\odot}$.  The bottom panel
shows the heliocentric radial velocity versus $l$ for the stream and
the satellite orbit.}
\end{figure}

Figure 6 shows that a fit to this realization gives fairly tight
confidence surfaces but only somewhat stronger than the phase-mixed,
$\lambda_{-10}=0.5$, $\vert\Delta\Theta\vert < 5^o$ fit given in
Figure 4 (lower left).  Although there are 4 times as many stars in
this fit, most of them are clumped near the satellite: this nullifies
the leverage of the additional stars.  The angular extent provides the
most leverage in fitting the projected orbit.

\begin{figure}
\plotone{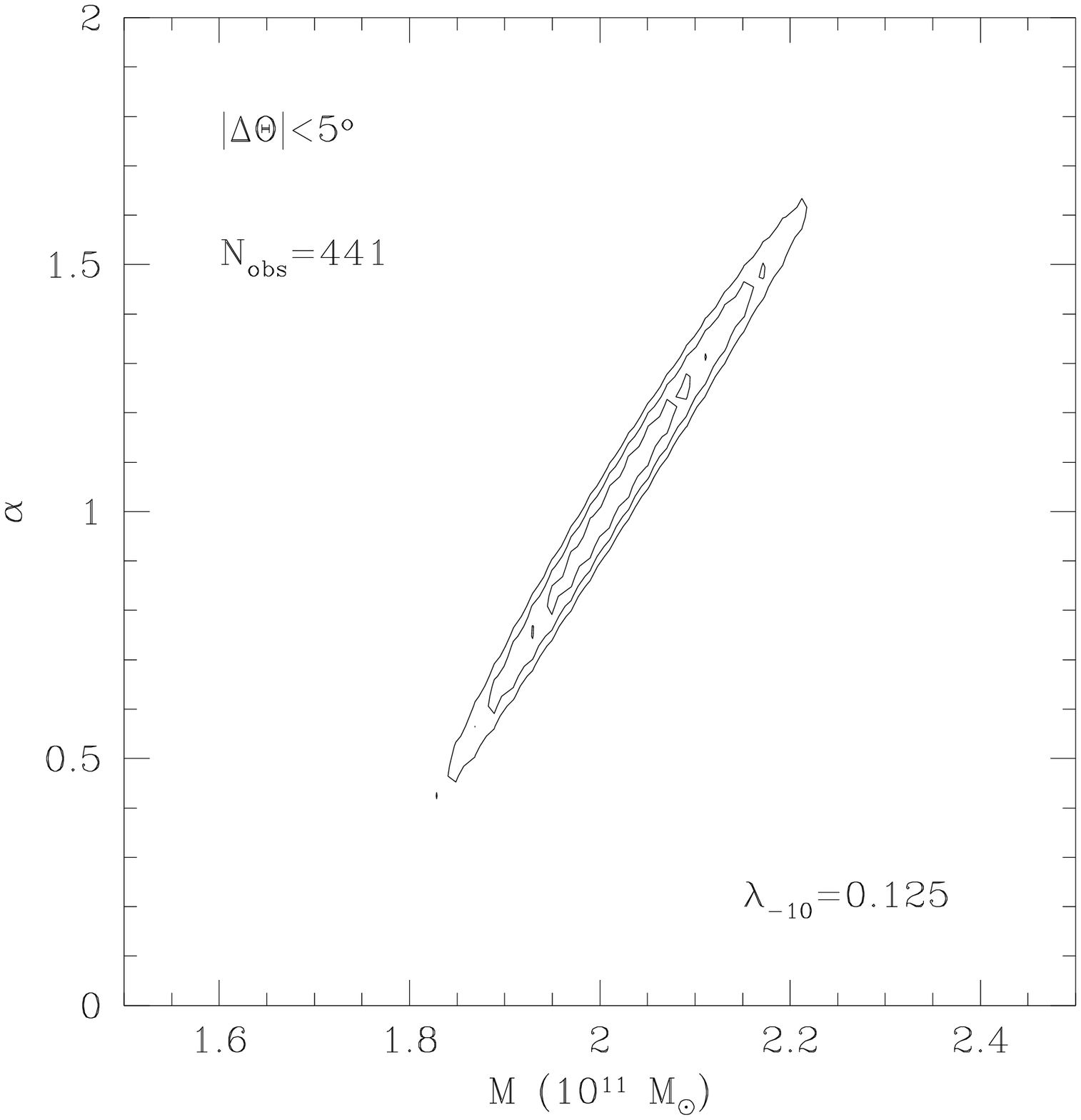}
\caption{Confidence intervals in $\alpha$ and $M_0$ for
$\vert\Delta\Theta\vert<5^o$ and $\lambda_{-10}=0.125$.  Contours show
the 99\%, 95\% and 68\% confidence contours.  The confidence surfaces
are fairly tight but are at nearly the same level as the surfaces for
$N_{obs}=60$, $\vert\Delta\Theta\vert < 5^o$ in the previous figure.}
\end{figure}

\subsubsection{The effect of proper motion uncertainties}
For illustrative purposes, we have so far assumed that available
measurements are perfect: i.e. that there are no uncertainties in
these quantities.  To be more realistic, we can apply the Bayesian
approach presented in \S\ref{sec:pm_sig} to account for measurement
uncertainties.  At present, proper motion uncertainties dominate,
especially for Pal 5: here we consider their effect on the estimated
mass.

Table 1 shows how confidences in the estimated value of $M_0$, denoted
$\bar M$, change when we include proper motion uncertainties in the
fits.  Proper motion uncertainties are defined relative to the mean
measured proper motion: {\it actual} refers to the uncertainties given
by Scholz et al. (1998).  The error bars $\sigma_{\bar M}$ define 95\%
confidence intervals about $\bar M$.

\begin{table*}
\caption{Estimated $M_0$ and errors with proper motion uncertainties}
\begin{tabular}{lllc}
\\ \hline
$\sigma_{pm}$ (rel.) & $\bar M (10^{11} M_{\odot})$ & 
$\sigma_{\bar M} (10^{11} M_{\odot})$ & $\Delta\Theta$ \\
\hline
$1\%$ & $2.0$ & $\pm 0.2$ & $5^o$ \\\\
$10\%$ & $1.7$ & $\pm{0.5 \atop 0.2}$ & $5^o$ \\\\
actual & $1.4$ & $\pm{0.3 \atop 0.3}$ & $5^o$ \\\\
actual & $1.8$ & $\pm{0.4 \atop 0.3}$ & $10^o$ \\\hline\\
\multispan4 error bars indicate 95\% confidence intervals
\end{tabular}
\end{table*}

As the table shows, 1\% proper motion uncertainties have little effect
on the fit: the mass estimate is unbiased and has tight, symmetric
confidence levels which are Gaussian. However, as we decrease the
proper motion accuracy, the fits degrade and estimates of $M_0$ appear
to become biased to lower values and confidences become highly skewed.
Qualitatively, the biasing results because, at lower $M_0$, there is a
broad range of $(\mu_{\alpha},\mu_{\delta})$ within the uncertainties
which give acceptable fits; at the true value, only a small range
gives an acceptable fit.  Therefore, in projection, it appears that
the lower value of $M_0$ is more likely.  The surfaces cut off rapidly
because the range of proper motions giving acceptable fits moves far
outside of the range of likely proper motions determined from the
observations.

The situation is rather bad for proper motion uncertainties comparable
to those which currently plague Pal 5: $\bar M$ is quite far from its
true value.  By extending the angular length of the observations, the
situation improves considerably because differences in orbit become
more pronounced over larger angles on the sky.  Ultimately, however,
the exact nature of this behavior depends on the particular cluster
under study and must be considered on a case-by-case basis.

\section{Discussion and conclusions}
\label{sec:discussion}
We have presented two methods for measuring the mass and potential of
the Galaxy using tidal streams from globular clusters.  The method
with full phase space information is closely related to previous work
by Lynden-Bell (1982), Kuhn (1993) and Johnston et al. (1998) and
clarifies the idea of using tidal streams as potentiometers.  It is
both a generalization of rotation curve measurements to non-circular
orbits and a dynamical statement of Gauss' law.  In principle, direct
measurements of the local gravitational acceleration and density field
can be obtained.  Of course, it should be emphasized that extremely
accurate data are required so current and near-future observations
will provide only a limited sample of objects to use with this method.

Analysis of the orbit-fitting method indicates that important
preliminary information can be obtained from incomplete phase-space
information using current observational capabilities.  However, we
must be careful to include measurement uncertainties in our analysis.
In particular, excessive proper motion uncertainties can introduce
biases into estimates.  In a broader context, we face the issue of
bias in any parametric analysis of the Galactic mass distribution,
regardless of measurement error: we must be careful not to ignore
systematic biases introduced by adopting any particular model for the
Galaxy.

The results presented above suggest that several good candidates in
the globular cluster population are available for halo mass
determinations at intermediate distances $R\sim 20\kpc$.  These
include Pal 5 as mentioned above; NGC 4147, NGC 5024 and NGC 5466 from
the Hipparcos sample; and possibly other clusters from ongoing plate
measurement programs (e.g. Dinescu et al 1999).  These clusters may
prove particularly interesting because of their large distance from
the Galactic plane: they could provide information on the mass
distribution in relatively uncharted territory.  Moreover, they
provide independent measurements of roughly the same region of the
Galaxy because of their spatial proximity.

The Hipparcos sample contains 15 clusters in all.  The clusters which
we have not mentioned lie closer to the disk and sun and, therefore,
have much better proper motion determinations and are easier to
observe.  In continuing the work presented here, we will develop
detailed models of tidal streams for these clusters in realistic,
axisymmetric models of the Milky Way (Kuijken \& Dubinski 1995).
These clusters will provide excellent candidates for probing the mass
distribution very close to the disk.  In particular, the astrometric
accuracy of the SIM and GAIA satellites within $20\kpc$ should allow
precise distance determinations to clusters and their streams.

There are clearly several uncertainties in the analysis presented
above.  The principal uncertainty lies in determining cluster mass
loss rates, which are difficult to model precisely.  A related
uncertainty is the stellar content of the stream.  On the one hand,
low mass clusters provide good candidates because they tend to lose
fair amounts of mass through relaxation and tidal heating.  However, a
large fractional mass loss from a low mass cluster is still a
relatively small number of stars.  Moreover, the stellar content of
the stream will play a strong role in its detectability.  Above we
adopted $\langle m\rangle=0.5 M_{\odot}$, which provides a fair number
of stars in the stream.  However, at $20 \kpc$, these stars would have
$I\sim 22$.  It would be extremely difficult-- if not impossible-- to
obtain $\km \sec^{-1}$ radial velocity accuracies for such faint
stars.  The best that we know of are Vogt et als (1995) radial
velocity study of $I=18-19$ giants in Leo II with Keck which required
$10$ minute exposures.  If, instead, the present mass function has
flattened substantially through dynamical evolution (e.g. Pal 5; Smith
et al. 1986), then many of the stream stars near the cluster may be
bright, higher mass stars.  This reduces the number of stream stars
for our adopted mass loss rates; but, if the mass loss rate is
somewhat higher, the situation may prove ideal.

Proper motions for these distant clusters also remain fairly uncertain
in spite of the success of Hipparcos.  There is presumably no hope of
improving stellar proper motion measurements before the next
generation of space-based, astrometric satellites.  However, we note
that, in the past, there has been some effort to detect water maser
emission from giants in globular clusters (Frail \& Beasley 1994).
Although unsuccessful, the possibility of improving proper motion
determinations using VLBI warrants renewed searches with deeper
detection limits.

The discussion and analysis presented above suggest that important
problems of the structure of the inner Galaxy may be fruitfully
addressed by searching for tidal streams from relatively nearby
globular clusters.  Moreover, with upcoming mission such as SIM
(scheduled for 2005) and GAIA (scheduled for 2009), it is essential to
establish groundwork for the larger scale studies advocated by
Johnston et al (1998) and Zhao et al (1998).  We expect that, with
present observational capabilities, important progress can be made by
focusing attention on the clusters discussed herein.  With these
missions still relatively far in the future, the progress made now can
help resolve important questions and serve as an invaluable guide.

\begin{acknowledgments}
We thank Bill van Altena and the referee, Kathryn Johnston, for
helpful discussion and acknowledge support from NSERC and the Fund for
Astrophysical Research.
\end{acknowledgments}

\newpage

\end{document}